# Mean Capacity of Spatially Semi-Correlated MIMO Fading Channel

Md. Abdul Latif Sarker and Moon Ho Lee, Sr. *Member, IEEE*

*Abstract*—This study investigates the mean capacity of multiple-input multiple-output (MIMO) systems for spatially semi-correlated flat fading channels. In reality, the capacity degrades dramatic due to the channel covariance (CC) when correlations exist at the transmitter or receiver or on both sides. Most existing works have so far considered the traditional channel covariance matrices that have not been entirely constructed. Thus, we propose an iterative channel covariance (ICC) matrix using a matrix splitting (MS) technique with a guaranteed zero correlations coefficient in the case of the downlink correlated MIMO channel, to maximize the mean capacity. Our numerical results show that the proposed ICC method achieves the maximum channel gains with high signal-to-noise ratio (SNR) scenarios.

## I. INTRODUCTION

MIMO wireless communications offer significant channel gains as well as improved diversity advantages and link robustness [1]. The practice of diversity and multiplexing techniques, the MIMO system exploits the spatial components of the wireless channel, such as optimal transmission, beamforming (BF), spatial multiplexing (SM), the and space-time or frequency code [2], which exploits the channel characteristics in different ways. For example, in the high level of spatially correlated MIMO channels, it is well-known that robust diversity-based schemes like BF should be employed. Elsewhere, a rich scattering SM method yielding high spectral efficiencies is more appropriate. In [3], different interesting correlation environments were employed to obtain the capacity mean and variance with equal power allocation. The most important technique was applied in [4-5] to obtain adaptive and asymptotic results for the capacity of a correlated channel.

Many studies [2, 4-5] have applied a different kind of adaptive algorithm using unlike construction of channel covariance matrices between the transmitter or receiver side correlations for maximizing the mean capacity of the correlated channel; Furthermore, expectations cannot be computed in closed-form [2, 5] when the channel state information (CSI) is not available at the transmitter side. Therefore, we propose an iterative covariance algorithm of a downlink correlated MIMO channel to assume the maximum mean capacity.

The results of the mean capacity of different types of channel realizations [6], for instance, unknown or known channel realization, are available at the transmitter or receiver sides; if the transmitter does not know the CSI, the power allocation matrix is proportional to the identity matrix, which is full-rank; otherwise, it may not be of full rank. In particular, the channel knowledge availability does not help to improve the mean capacity when the average SNR is extremely high [6]. However, we consider unknown channel knowledge under a high SNR regime to compute the channel gain easily.

This remainder of this paper is arranged as follows. First, we investigate the channel model and problem formulation in Section II. Then, we propose an iterative channel covariance matrix and maximum mean capacity in Section III. Finally, the numerical results and conclusions are presented in Sections IV and V, respectively.

## II. CHANNEL MODEL AND PROBLEM FORMULATION

We consider a narrowband flat fading MIMO downlink system with $N_t$ transmitting and $N_r$ receiving antenna, which can be modeled as:

$$\mathbf{y} = \sqrt{E_s}\mathbf{H}\mathbf{x} + \mathbf{z} \qquad (1)$$

where $\mathbf{y} \in \mathbb{C}^{N_r \times 1}$ and $\mathbf{x} \in \mathbb{C}^{N_t \times 1}$ are the received and transmitted signal vectors, respectively, $E_s$ is the average transmitted energy per antenna, $\mathbf{z} \in \mathbb{C}^{N_r \times 1}$ is a complex additive white Gaussian noise (AWGN) vector with covariance matrix $\mathbb{E}\{\mathbf{z}\mathbf{z}^H\} = N_0\mathbf{I}_{N_r}$, the operator $(\cdot)^H$ is called a Hermitian, $\mathbf{I}_{N_r}$ is the $N_r \times N_r$ identity matrix, and $\mathbf{H} \in \mathbb{C}^{N_r \times N_t}$ is the assumed $N_r \times N_t$ complex channel matrix of independent and identically distributed (i.i.d) circular Gaussian variables with zero mean and unit total variance, as is in the independent and identical fading channel. The SNR is defined as $\gamma_0 = E_s / N_0$.

We consider an input covariance matrix $\mathbf{Q}_{\mathbf{xx}} = \mathbb{E}\{\mathbf{x}\mathbf{x}^H\} = E_s\mathbf{I}_{N_t} / N_t$ as in [3] for spatial multiplexing

This work was supported in part by "MEST 2015R1A2A1A05000977, NRF, South Korea".

Md. Abdul Latif Sarker is with the Chonbuk National University, Jeonju, 54896 Korea (e-mail: latifsarker@jbnu.ac.kr).
Moon Ho Lee, was with Chonbuk National University, Jeonju-si, 54896 Korea. He is now with the Department of Electronics and Information Engineering, Conbuk National University, Jeonju-si, 54896 Korea (e-mail: moonho@jbnu.ac.kr).

(SM) transmission with equal power allocation across the $N_t$ transmit antennas; $\mathbf{I}_{N_t}$ is the $N_t \times N_t$ identity matrix. Thus, the mean capacity of the MIMO system without using CSI at the transmitter side is given by [1, 3-4]:

$$\bar{\mathbf{C}} = \mathbb{E}\left\{\log_2 \det\left(I_{N_r} + \frac{\gamma_0}{N_t}\mathbf{H}\mathbf{H}^H\right)\right\} bps/Hz \quad (2)$$

In the presence of correlation in the MIMO fading channel, most usual models presume that the fading is induced by separate physical processes at the transmitter [3,5], which assume that no correlation exists between the receiver antennas. This leads to the channel matrix being modeled as $\mathbf{H} = \mathbf{H}_w \mathbf{R}_t^{1/2}$, where the entries of $\mathbf{H}_w$ are $N_r \times N_t$ independent and identically distributed (i.i.d) Rayleigh fading channel and $\mathbf{R}_t$ is a $N_t \times N_t$ transmitted covariance matrix [2–5] that reflects the correlations between the transmitting antennas. Using the correlated channel model in (2), the mean capacity is given as [4-5]:

$$\bar{\mathbf{C}}(\mathbf{R}_t) = \mathbb{E}\left\{\log_2 \det\left(I_{N_r} + \frac{\gamma_0}{N_t}\mathbf{H}_w \mathbf{R}_t \mathbf{H}_w^H\right)\right\} \quad (3)$$

We observe in (3) that the transmitter covariance matrix $\mathbf{R}_t$ makes an approximately Toeplitz assumption in [7-8], which restricts the actual configuration of a uniform linear array (ULA) to decompose large singular value decomposition (SVD) systems in [2, 4-5, 9]. In addition, the expectations cannot be computed in the closed-form of (3). However, the amount of channel capacity reduction is dramatic due to the correlations between the transmitting and receiving antennas. Thus, we next construct an ICC algorithm in the Section III.

### III. PROPOSED ICC MATRIX AND MAXIMUM MEAN CAPACITY

We now construct an iterative channel covariance (ICC) algorithm for obtaining the maximum mean capacity. To generalize the channel covariance matrix, we consider $N_t = N_r = N$, and $\mathbf{R}$ is full rank in high SNR regimes; then, we formulate *Theorem 1* as follows:

*Theorem 1:* The Toeplitz assumption of any $N \times N$ covariance matrix $\mathbf{R}$ splits and possesses a sum of two $N \times N$ equivalent right circulant matrices, and is given as [10-11]:

$$\mathbf{R} = \mathbf{A} + \mathbf{B} \quad (4)$$

where $\mathbf{A} = circ[a_0 \ a_1 \ \ldots \ a_{N-1}]$ is a $N \times N$ right circulant and $\mathbf{B} = [b_{i-j}]_{j,i=1}^{N}$ is a $N \times N$ skew right circulant matrix where $b_{-j} = -b_{N-j}$ for $j = 1, \ldots N-1$.

**Proof** of *Theorem 1*: see Appendix.

However, finally, we would like to analyze the maximum mean capacity depending on the variations of the channel covariance matrix $\mathbf{R}$. To improve (3), we can construct an iterative channel covariance matrix $\mathbf{R}(\alpha)$ given by [10]:

$$\mathbf{R}(\alpha) = (\alpha\mathbf{I} + \mathbf{B})^{-1}(\alpha\mathbf{I} - \mathbf{A})(\alpha\mathbf{I} + \mathbf{A})^{-1}(\alpha\mathbf{I} + \mathbf{B}) \quad (5)$$

where the spectral radius $\rho(\mathbf{R}(\alpha))$ is bounded by

$$\sigma(\alpha) \equiv \max_{\lambda_j \in \lambda(\varepsilon_1)} \frac{|\alpha - \lambda_j|}{|\alpha + \lambda_j|} \cdot \max_{\mu_j \in \mu(\varepsilon_2)} \frac{|\alpha - \mu_j|}{|\alpha + \mu_j|}, \quad (6)$$

where $\alpha > 0$ is a positive constant, and $\lambda_j$ and $\mu_j$ are the eigenvalues of $\mathbf{A}$ and $\mathbf{B}$, respectively.

Thus, it holds that

$$\rho(\mathbf{R}(\alpha)) \leq \sigma(\alpha) < 1, \ \forall_{\alpha > 0}. \quad (7)$$

By substituting (5) into (3), the maximum mean capacity at the transmitter side correlation is given by:

$$\bar{\mathbf{C}}(\mathbf{R}_t(\alpha)) = \max_{\alpha > 0} \mathbb{E}\left\{\log_2 \det\left(I_{N_r} + \frac{\gamma_0}{N_t}\mathbf{H}_w \mathbf{R}_t(\alpha)\mathbf{H}_w^H\right)\right\} \quad (8)$$

where the maximum parameter $\alpha_{max}$ is chosen such that the convergence ensures the zero correlation coefficient of the transmitted covariance matrix $\mathbf{R}_t(\alpha_{max})$, which should be the identity matrix i.e., $\mathbf{R}_t(\alpha_{max}) = \mathbf{I}_t(\alpha_{max})$: Tables I and II.

### IV. NUMERICAL RESULTS

We now provide numerical results for the conventional CC methods and the proposed ICC methods for a correlated fading channel compared to an uncorrelated independent and identically distributed (i.i.d) Rayleigh fading channel. In view of *Theorem 1*, we generate $\mathbf{R}$ from [5]. Then, we split $\mathbf{R}$ as (4) and regenerate $\mathbf{A}$ and $\mathbf{B}$ as in (16) and (17), shown in the Appendix. After that, we apply the proposed ICC algorithm in (8). In particular, we focus on the proposed ICC algorithm (Table I) using the different $\alpha$ values shown in Table II.

TABLE I
PROPOSED ICC ALGORITHM

| |
| --- |
| 1: Initialization: $\alpha > 0$, a positive constant. |
| 2: For $\alpha = 1$ to $\alpha_{max}$ |
| $\mathbf{R}_t(\alpha) = (\alpha\mathbf{I} + \mathbf{B}_t)^{-1}(\alpha\mathbf{I} + \mathbf{A}_t)(\alpha\mathbf{I} + \mathbf{A}_t)^{-1}(\alpha\mathbf{I} + \mathbf{B}_t)$. |
| 3: Finalize: |
| If $\alpha \neq \alpha_{max}$, then |
| $\mathbf{R}_t(\alpha_{max}) \neq \mathbf{I}_t(\alpha_{max})$, Still correlations are remaining i.e. $\rho > 0$. |
| 4: Return to new convergence. |
| If $\alpha = \alpha_{max}$, then |
| $\mathbf{R}_t(\alpha_{max}) = \mathbf{I}_t(\alpha_{max})$, no correlations i.e. $\rho = 0$. |
| 5: Stop convergence (end). |

Fig. 1 shows the mean capacities of two different channel environments versus cumulative distribution function (CDF) at 30 dBs SNR when $N_t = N_r = 4$. In the figure, we consider different $\alpha$ values to obtain the maximum mean capacities when CSI is not available at the transmitter side.

Fig. 2 depicts the mean capacities of two different channel approaches versus SNR. We observe that a capacity of 2.9 bps/Hz is lost in Fig. 1 and Fig. 2, due to the channel correlations at 30 dBs SNR. In the observations, when $\alpha$ values are increased, the average channel gain is also increased. For instance, when $\alpha = 5$ and 10, the channel average gains are 0.61 bps/Hz and 1.75 bps/Hz, respectively which is shown in Fig. 1(a) and Fig. 2(b). Similarly, when $\alpha = 20$ and 30, the channel average gains are 2.4 bps/Hz and 2.65 bps/Hz, as shown in Fig. 1(b) and Fig. 2(b), respectively, which are closed to uncorrelated (i.i.d) channel gains. Thus, we can ensure a zero correlation coefficient (Table II), in order to obtain 100% of the channel average gains as uncorrelated (i.i.d) Rayleigh fading

TABLE II
BOUNDARY OF SPECTRAL RADIUS IN (7)

| $\alpha > 0$ | $\sigma(\alpha)$ | $\rho(\mathbf{R}(\alpha))$ | Status |
|---|---|---|---|
| 5 | 0.6711 | 0.6901 | |
| 10 | 0.8112 | 0.8262 | |
| 20 | 0.8982 | 0.9062 | |
| 30 | 0.9303 | 0.9364 | |
| 50 | 0.9573 | 0.9611 | Correlations exist i.e., |
| 100 | 0.9783 | 0.9863 | $\rho > 0$. |
| 200 | 0.9891 | 0.9901 | |
| 600 | 0.9963 | 0.9967 | |
| 1,000 | 0.9978 | 0.9980 | |
| 20,000 | 0.9999 | 0.9999 | |
| 40,000 | 0.9999 | 1.0000 | Zero correlations, i.e., |
| 50,000 | 1.0000 | 1.0000 | $\rho = 0$ and |
| 60,000 | 1.0000 | 1.0000 | $\mathbf{R}(\alpha_{max}) = \mathbf{I}(\alpha_{max})$ |

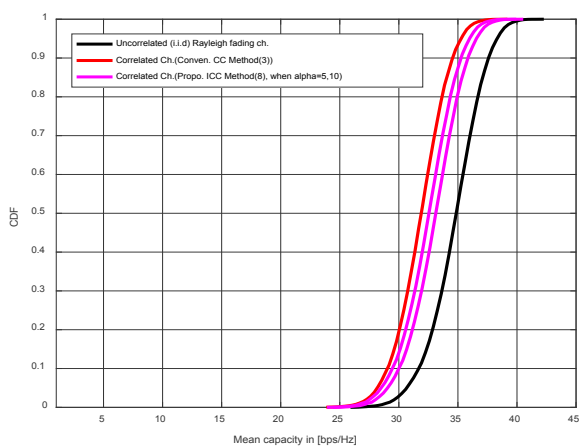

(a) CDF vs. Mean capacity when $\alpha = 2$, and 10.

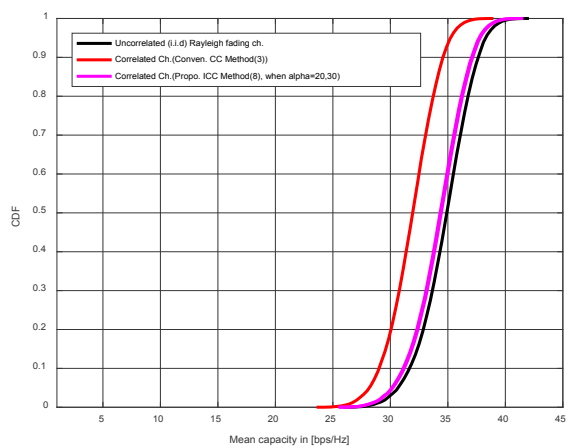

(b) CDF vs. Mean capacity when $\alpha = 20$, and 30

Fig. 1. CDF vs. mean capacity for the different $\alpha$ values at 30 dBs SNR.

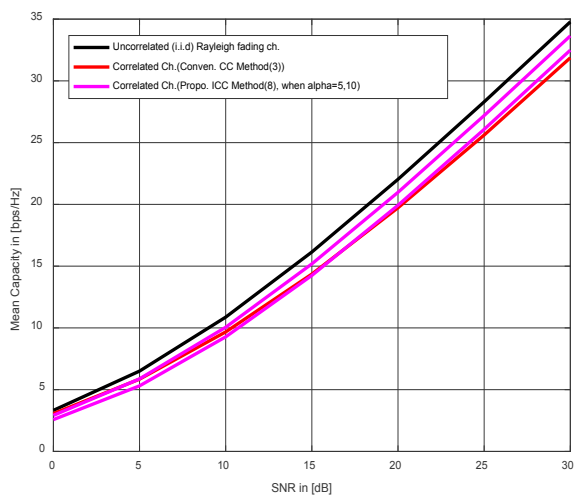

(a) Mean Capacity when $\alpha = 2$, and 10.

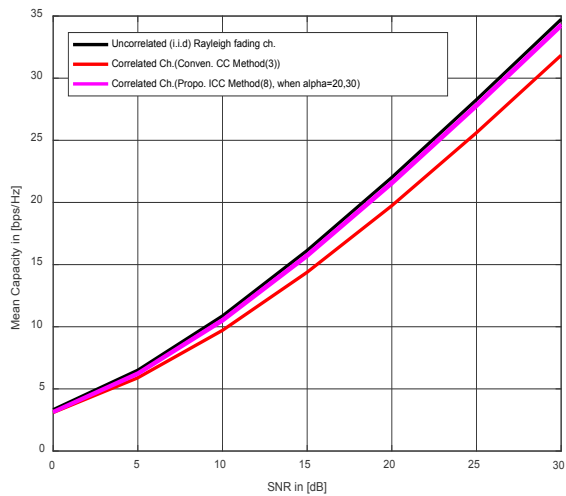

(b) Mean Capacity when $\alpha = 20$, and 30.

Fig. 2. Mean capacity for the different $\alpha$ values at 30 dBs SNR.

channel when the parameter $\alpha$ is a maximum ($\alpha_{max}$) in (8).

## V. CONCLUSION

To sum up, we have investigated the zero correlation coefficient of the transmitted covariance matrix using our ICC algorithm to maximize the mean capacity. Many existing studies concern the error covariance matrix convergence problems of correlated interference MIMO channels. We may further deal with the results presented here in the future.

APPENDIX

*Proof of Theorem 1:* Let the complex channel covariance matrix $\mathbf{R} = \mathbb{E}\{\mathbf{H}^H \mathbf{H}\}$ or $\mathbf{R} = \mathbb{E}\{\mathbf{H}\mathbf{H}^H\}$ be a $N \times N$ Toeplitz matrix given by [8, 10-11]:

$$\mathbf{R} = \begin{bmatrix} h_0 & h_{-1} & h_{-2} & \cdots & h_{-N+1} \\ h_1 & h_0 & h_1 & & \\ h_2 & h_1 & h_0 & & \vdots \\ \vdots & & & \ddots & \\ h_{N-1} & & \cdots & & h_0 \end{bmatrix}. \quad (9)$$

The corresponding elements of the error covariance matrix $\mathbf{R}$ lead to a system for $\mathbf{A}$ and $\mathbf{B}$ as follows [10]:

$$h_{-j} = a_j + b_j, \quad j = 0,1,\ldots,N-1 \quad (10)$$

$$h_{N-j} = a_j - b_j, \quad j = 1,2,\ldots,N-1 \quad (11)$$

We obtain from (9), $a_0 = h_0/2$ and $b_0 = h_0/2$ when $j = 0$ and $a_0 = b_0$. Using (10)–(11) for $j = 1,2,\ldots,N-1$ we have,

$$a_j = \frac{h_{-j} + h_{N-j}}{2} \text{ and } b_j = \frac{h_{-j} - h_{N-j}}{2}. \quad (12)$$

Thus, we can calculate $\mathbf{A}$ and $\mathbf{B}$ using (10)–(12):

$$\mathbf{A} = \begin{bmatrix} a_0 & a_1 & a_2 & \cdots & a_{N-1} \\ a_{N-1} & a_0 & a_1 & & \\ a_{N-2} & a_{N-1} & a_0 & & \vdots \\ \vdots & & & \ddots & \\ a_1 & & \cdots & & a_0 \end{bmatrix}, \quad (13)$$

and

$$\mathbf{B} = \begin{bmatrix} b_0 & b_1 & b_2 & \cdots & b_{N-1} \\ -b_{N-1} & b_0 & b_1 & & \\ \vdots & -b_{N-1} & b_0 & & \vdots \\ \vdots & & & \ddots & \\ -b_1 & & \cdots & & b_0 \end{bmatrix} \quad (14)$$

Thus, we observe that (13) and (14) are obviously circulant and skew circulant matrices.

*Example 1:* Let $N_t = 4$; the complex channel covariance matrix $\mathbf{R}_t$ at the transmitter side correlations of order 4 is given by [5]:

$$\mathbf{R}_t = \begin{bmatrix} 1 & -0.3581-0.4435i & 0.1700+0.0034i & -0.2841+0.0581i \\ -0.3581+0.4435i & 1 & -0.3581-0.4435i & 0.1700+0.0034i \\ 0.1700-0.0034i & -0.3581+0.4435i & 1 & -0.3581-0.4435i \\ -0.2841-0.0581i & 0.1700-0.0034i & -0.3581+0.4435i & 1 \end{bmatrix} \quad (15)$$

Applying (10)–(12) into (15), we can calculate $a_0 = 0.5000$, $a_1 = -0.3211 - 0.1927i$, $a_2 = 0.1700$, and $a_3 = -0.3211 + 0.2508i$, respectively. Similarly, we can also calculate $b_0 = 0.5000$, $b_1 = -0.0370 - 0.2508i$, $b_2 = 0 + 0.0034i$, and $b_3 = 0.0370 - 0.1927i$, respectively. Substituting all of these values into (13) and (14), we have

$$\mathbf{A}_t = \begin{bmatrix} 0.5000 & -0.3211-0.1927i & 0.1700 & -0.3211+0.2508i \\ -0.3211+0.2508i & 0.5000 & -0.3211-0.1927i & 0.1700 \\ 0.1700 & -0.3211+0.2508i & 0.5000 & -0.3211-0.1927i \\ -0.3211-0.1927i & 0.1700 & -0.3211+0.2508i & 0.5000 \end{bmatrix} \quad (16)$$

and

$$\mathbf{B}_t = \begin{bmatrix} 0.5000 & -0.0370-0.2508i & 0+0.0034i & 0.0370-0.1927i \\ -0.0370+0.1927i & 0.5000 & -0.0370-0.2508i & 0+0.0034i \\ 0-0.0034i & -0.0370+0.1927i & 0.5000 & -0.0370-0.2508i \\ 0.0370+0.1927i & 0-0.0034i & -0.0370+0.1927i & 0.5000 \end{bmatrix} \quad (17)$$